\documentclass[groupedaddress,endfloats*]{revtex4-1}
\usepackage{epsfig}

\usepackage{graphics}
\usepackage[latin1]{inputenc}
\usepackage[english]{babel}
\usepackage{gensymb}
\usepackage{color}

\begin{document}
\title{Controlling boron redistribution in CoFeB/MgO magnetic tunnel junctions during annealing by variation of cap layer materials and MgO deposition methods}

\author{Henning Schuhmann}
\author{Michael Seibt}
\email{mseibt@gwdg.de}
\homepage{http://www.uni-goettingen.de/en/101067.html}
\affiliation{IV. Physikalisches Institut, Georg-August-Universität Göttingen, Friedrich-Hund-Platz 1, D-37077 Göttingen, Germany}
\author{Volker Drewello, Andy Thomas}
\affiliation{Thin Films and Physics of Nano Structures, Department of Physics, Bielefeld University, 33501 Bielefeld, Germany}
\author{Vladyslav Zbarsky, Marvin Walter, Markus Münzenberg}
\affiliation{I. Physikalisches Institut, Georg-August-Universität Göttingen, Friedrich-Hund-Platz 1, D-37077 Göttingen, Germany}
\affiliation{Institut für Physik, Universität Greifswald, Felix-Hausdorff-Str. 6, D-17489 Greifswald, Germany}

\begin{abstract}
Magnetic tunnel junctions with crystalline MgO tunnel barrier and amorphous CoFeB electrodes received much attention due to their high tunnel magneto resistance ratio at room temperature. One important parameter for achieving high tunnel magneto resistance ratios is to control the boron diffusion from the electrodes  especially during post growth annealing. By high resolution transmission electron microscopy and electron energy loss spectroscopy techniques we show that  the cap layer material adjacent to the electrodes and the MgO deposition method are crucial to control boron redistribution. It is pointed out, that Ta cap layers acts as sinks for boron during annealing in contrast to Ru layers.  Furthermore,  radio frequency sputtered MgO tunneling barriers contain a rather high concentraion of boron in trigonal [BO$_3$]$^{3-}$ - environment after annealing in contrast to electron beam evaporated MgO which is virtually free from any boron.  Our data further indicate that neither boron nor oxygen-vacancy-related gap states in the bulk of MgO barriers affect spin polarized transport for tunnel magneto resistance ratios at the level of 200\%.
\end{abstract}\maketitle

In recent years, magnetic tunnel junctions (MTJ) with crystalline MgO tunnel barrier and amorphous CoFeB electrodes received much attention. Due to their high tunneling magnetoresistance (TMR) ratio at room temperature they are a promising candidate for magnetic random access memory or as sensor elements for harddrive read heads \cite{Miao2011}. 
For industrial applications, the use of amorphous ferromagnetic electrodes in MTJ is favorable due to their better growth flexibility compared to crystalline electrodes. {After annealing at temperatures up to 600\degree C high TMR was achieved  \cite{Karthik2012, Thomas2008, Schafers2009}. This effect is related to partial crystallization of the CoFeB into a bcc crystal structure using the MgO as a template \cite{Eilers2009b}. The resulting enhanced conduction through the MTJ can be understood in terms of coherent tunneling process through MgO $\Delta_1$-band \cite{Burton2006,Heiliger2007}.  }

 The (partial) crystallization of a-CoFeB is intimately connected to boron out-diffusion \cite{Schreiber2011}. Hence, the ability of layers adjacent to a-CoFeB to act as sinks or diffusion barriers for boron is crucial to achieve high TMR during annealing in a controlled way. Besides the MgO barrier sandwiched between the a-CoFeB electrodes, cap layers of the MTJ stack influence boron diffusion and crystallization of the ferromagnetic electrode~\cite{Schreiber2011}.

Furthermore, the MgO deposition method seems to have a big influence on the segregation of boron at the MgO / CoFe interface. Earlier studies showed a significant amount of B and $\text{BO}_x$ at the interface of radio frequency (rf) (magnetron)sputtered \cite{Cha2009, Karthik2009, Schreiber2011} barriers, but not in electron beam (eb) evaporated MgO barriers \cite{Kurt2010,Cha2007}. Due to the high affinity of boron to oxygen \cite{Kurt2010}, the formation of $\text{BO}_x$ at the interface is believed to be caused by the incorporation of excess oxygen from the ambient vacuum during rf-sputtering \cite{Schreiber2011}.

In this study, the effect of the most promising cap layer materials \cite{Schreiber2011}, Ta and Ru, on boron redistribution was studied by electron energy loss spectroscopy (EELS). The related crystallization process of a-CoFeB grown on high quality eb-evaporated MgO was examined by high-resolution transmission electron microscopy (HRTEM).
Furthermore, the influence of the MgO deposition method used on boron redistribution was studied on two functional MTJ.
 It turned out that Ta cap layer acts as a sink for boron and do not as a template for a-CoFeB crystallization. In addition, our results provide evidence that unlike rf-sputtered MgO tunnel barriers, eb-evaporated MgO barriers do not contain boron after post-growth annealing.

In order to study the effect of capping layer materials, a model system consisting of a MgO substrate covered by an eb-deposited MgO  (thickness: 5 nm) buffer layer, $\text{Co}_{20}\text{Fe}_{60}\text{B}_{20}$ (5 nm or 100 nm) and a cap layer was grown under UHV conditions at a base pressure of $3 \cdotp10^{-10}$ mbar. A 10 nm Ru layer or a 10 nm Ta + 3 nm Ru layer, respectively, act as the cap layer. The role of the Ru on top of the Ta layer is merely to prevent the layer stack from oxidation so that this assembly will subsequently be referred to as \grqq Ta cap layer\grqq. 
Both cap layers were grown on a 5 nm and 100 nm CoFeB layer and annealed at 450 °C for one hour. 

The effect of the MgO barrier deposition method was investigated by preparing two functional MTJ with slightly different layer stacks.
The MTJ with an eb-evaporated MgO tunnel barrier used in this study was grown on a $\text{SiO}_2$ wafer with a layer stack of Ta(5) / CoFeB(2.5) / MgO(2.3) / CoFeB(5.4) / Ta(5) / Ru(3) (thickness in nm) and was annealed at 375 °C for one hour.
The rf-sputtered MTJ was also grown on a $\text{SiO}_2$ wafer, but in contrast to the previous sample with a slightly different layer stack of Ta(5) / Ru(30) / Ta(10) / Ru(10) / CoFeB(4) / MgO(2.1) / CoFeB(1.5) / Ta(5) / Ru(20) and an annealing time of one hour at 400 °C. The base pressure during MgO growth was approximately $1 \cdotp10^{-9}$ mbar.
Both samples showed a very similar TMR of around 200 \% measured with 10 mV bias voltage.

Cross-section TEM lamella preparation was done using a FEI Nova NanoLab 600 Focused Ion Beam with a final polishing step under 5 kV. Subsequent Ar ion milling in a Gatan Precision Ion Polishing System at a voltage of 500V was used to minimize beam damage. TEM observations and EELS studies were carried out using an imaging aberration corrected FEI Titan 80-300 ETEM G2 operated at 300kV, equipped with a monochromator and a Gatan imaging filter Quantum ER965. EELS spectra were taken in scanning TEM (STEM) mode with a probe size of 1.5 \AA, a convergence semiangle of $\sim 9.5$ mrad and a spectrometer dispersion of 0.25 eV/channel resulting in an energy resolution of 1.5 eV. To minimize radiation damage, EELS spectra were acquired and integrated over multiple points parallel to the MgO/CoFe interface.

\begin{figure}
\centerline{
 \includegraphics[width=0.95\columnwidth]{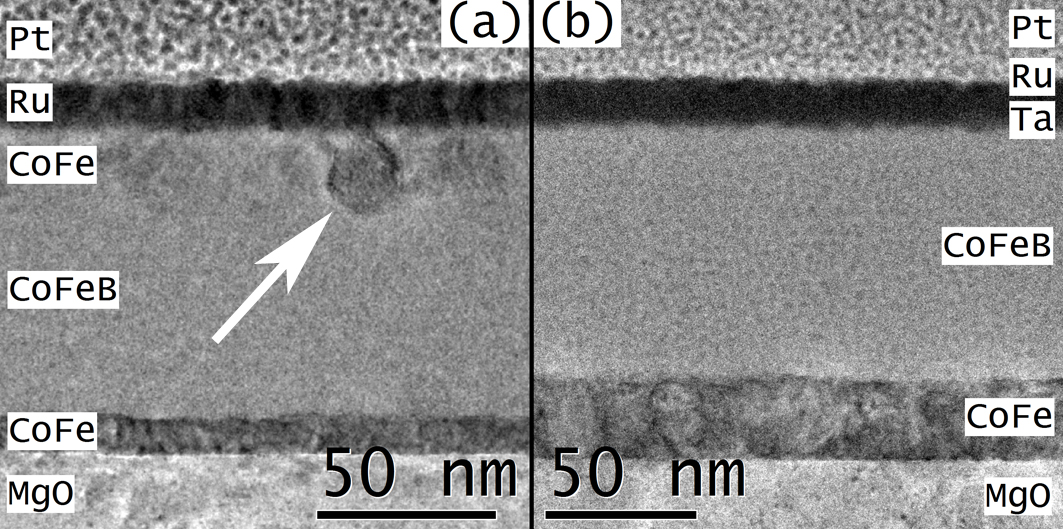}}
\caption{TEM images of samples with a 100 nm thick CoFeB layer: Ru cap layer (left) and Ta+Ru cap layer (right). The lower crystalline CoFe layer is clearly visible in both samples. A second polycrystalline CoFe region is present in the Ru cap sample at the interface to the cap layer (see arrow).}
\label{fig:1}
\end{figure}

{ Fig.~\ref{fig:1}a and b compare crystallization of 100nm thick a-CoFeB layers for Ru and Ta cap layers, respectively. Crystalline CoFe layers epitaxially grown on MgO are observed in both cases with a thickness of 13nm (Ru) and 25nm (Ta) proving faster crystallization in the latter case. Furthermore, polycrystalline CoFe has grown beneath the Ru cap (see arrow in Fig.~\ref{fig:1}a) implying the polycrystalline hcp Ru acting as a template for CoFeB crystallization in contrast to the nanocrystalline Ta.}

\begin{figure}
\centerline{
 \includegraphics[width=0.95\columnwidth]{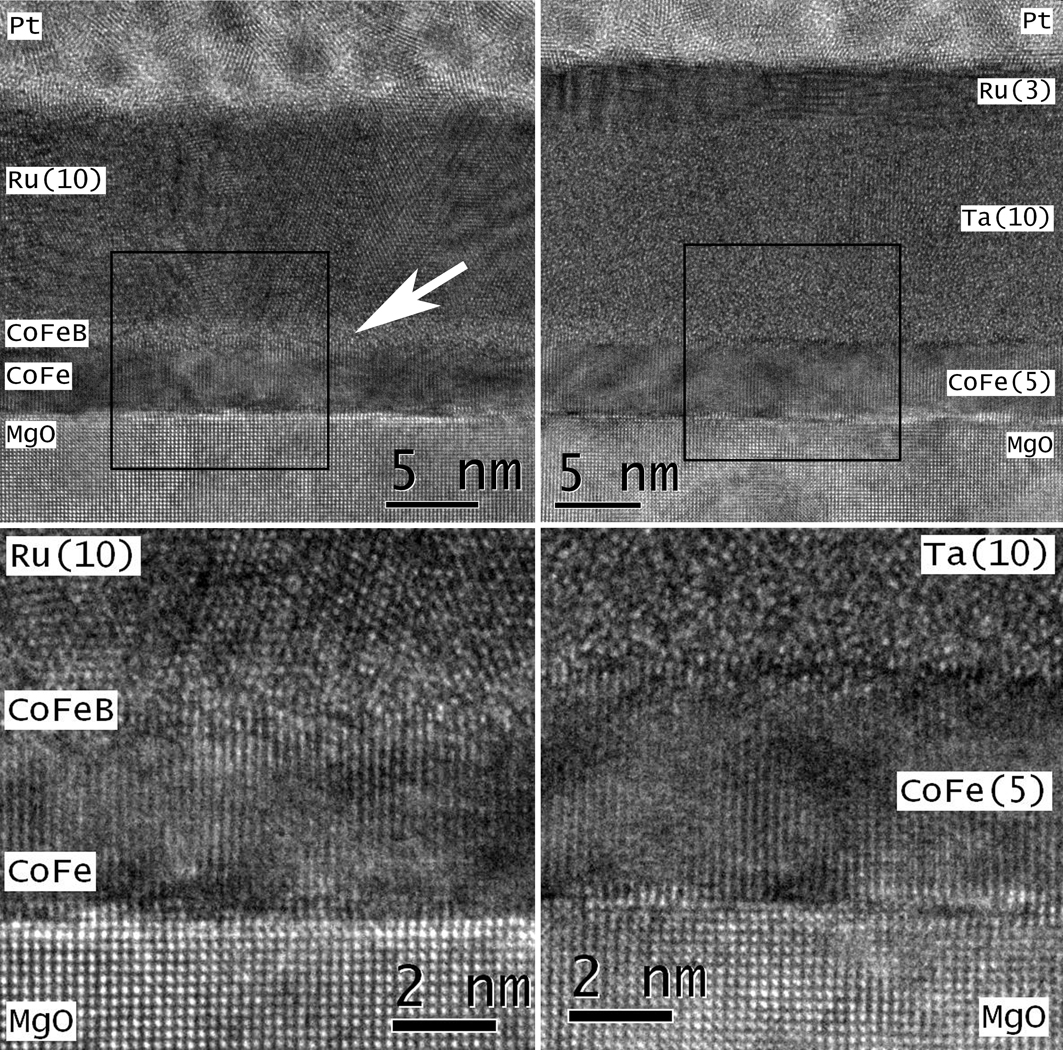}}
\caption{{HRTEM images of samples with a 5 nm thick CoFeB layer with Ru cap layer (left) and Ta+Ru cap layer (right). The a-CoFeB is only partially crystallized leaving an  about 1nm thick amorphous layer beneath the Ru cap (left) whereas full crystallization is observed for the Ta cap layer (right). Boxes indicate in a) and b) indicate  position of details shown in c) and d), respectively. Please note the somewhat brighter appearance of the Ta layer close to the crystalline CoFe.}}
\label{fig:2}
\end{figure}

Similarly, a-CoFeB crystallized into bcc CoFe at the interface to the MgO in samples with 5nm CoFeB layers (Fig. \ref{fig:2}). A thin a-CoFeB layer is left in case of the Ru cap (Fig.~\ref{fig:2}a and c) 
whereas full crystallization is observed in case of the Ta cap (Fig.~\ref{fig:2}b and d) which again shows the faster crystallization in the latter case. It should be noted that according to the results obtained for the thick layer samples full crystallization is expected irrespective of the cap layer material which will be discussed below in terms of results on boron redistribution. 

\begin{figure}
\centerline{
 \includegraphics[width=1\columnwidth]{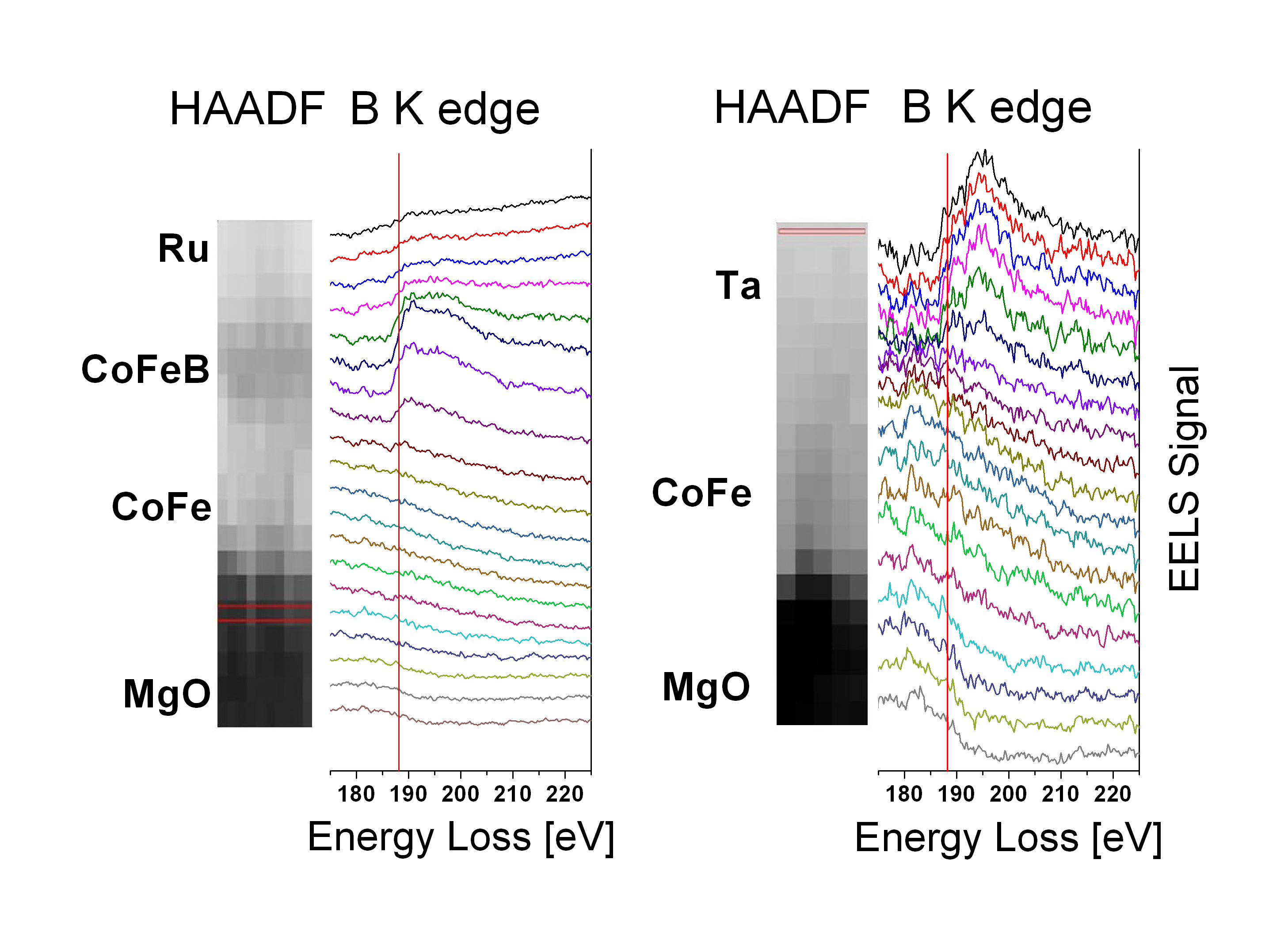}}
\caption{(color online) EEL spectra of the B-K edge and O-K edge from the MgO into the Ru cap. The absence of B and $\text{BO}_x$ at the MgO and its interface is clearly visible from these spectra. The boron is only present in the amorphous interlayer between the Ru cap and the CoFe.}
\label{fig:3}
\end{figure}

{ Fig.~\ref{fig:3} summarizes representative EELS linescans acquired from the MgO layer across the CoFe layer into the Ru (left) or Ta (right) cap for the energy region of the boron absorption edge around 188 eV. For both cap layer materials, no boron-related signal is visible inside the MgO, at the interface between MgO and CoFe, and in the crystallized CoFe. This behavior is reproducible for all samples. For the Ru cap, virtually all boron is localized in the amorphous interlayer which appears as a slightly darker layer in the HAADF signal. In particular, no significant boron signal is obtained inside the Ru cap itself. For the Ta cap layer sample, boron is exclusively obtained in the Ta layer indicating complete out-diffusion from the CoFeB during crystallization.  Its signal steeply decreases with increasing distance to the Ta/CoFe interface. Hence, we might conclude that the brighter appearance of the Ta close to the CoFe (see Fig.~\ref{fig:2}b and d) is related to its boron content. It should be noted that no boron is found in any crystallized CoFe, at the CoFe/MgO interface, at the Ru cap interface of the thick CoFeB sample, and, in particular, in the e-beam deposited MgO buffer layers.} { In conclusion, the above results provide evidence that Ru layers act as a boron diffusion barrier which reduces the crystallization velocity of adjacent a-CoFeB layers. This effect is more severe for thin a-CoFeB layers indicating that the crystallization velocity is limited by the increase of boron concentration in the remaining a-CoFeB rather than boron segregation at the interface between crystalline CoFe and a-CoFeB.}

\begin{figure}
\centerline{
 \includegraphics[width=0.9\columnwidth]{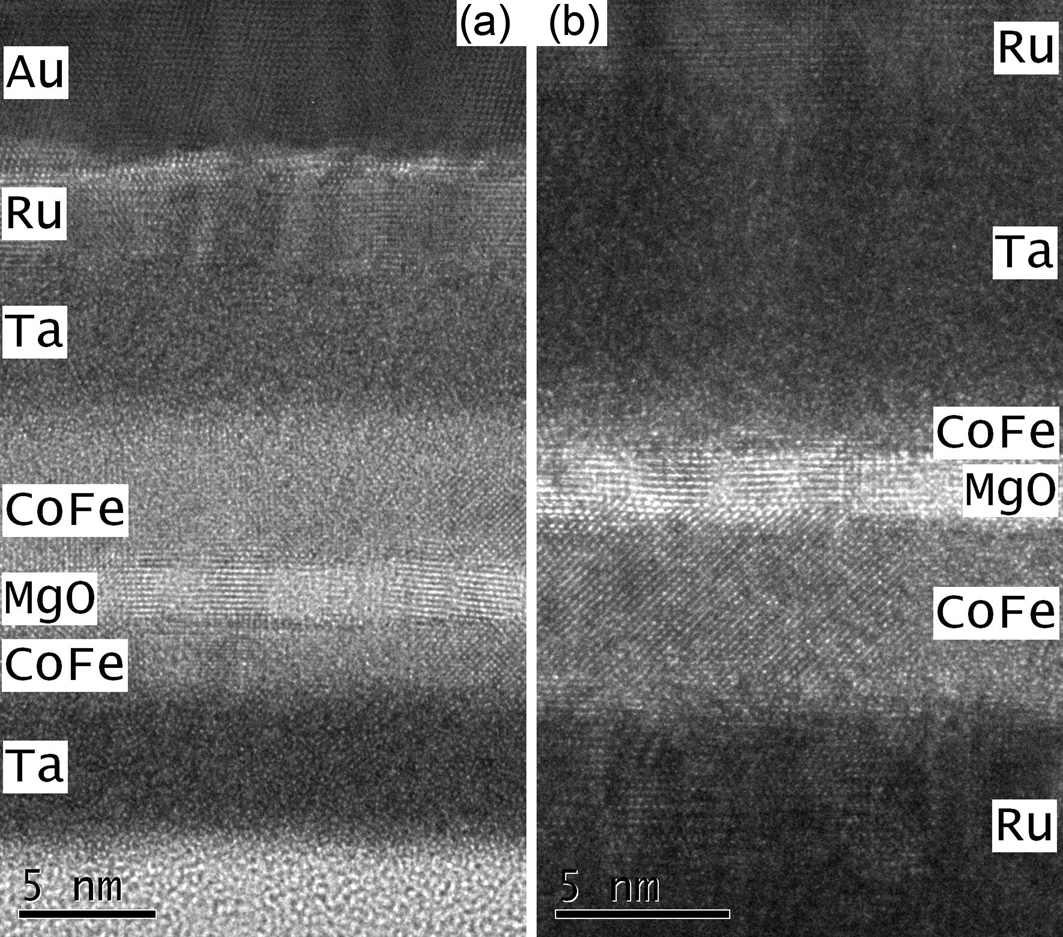}}
\caption{HRTEM micrographs of a MTJ with eb-evaporated MgO barrier (a) and rf-sputtered barrier (b). Both show a crystalline MgO barrier and crystallized electrodes.}
\label{fig:4}
\end{figure}

{ Besides the choice of cap layer material, the preparation method of the MgO barrier is an important issue when functional MTJ are considered. We prepared TEM lamellas from MTJ stacks containing eb-evaporated and rf-sputtered and post-annealed MgO barriers as described in detail above. Fig. \ref{fig:4} shows homogeneous and crystalline MgO layers with crystallized CoFeB electrodes and sharp interfaces for both techniques. } As in the Ru cap layer sample of the model system (Fig. \ref{fig:2}a), a small amorphous region with slightly brighter contrast is visible between top CoFe and the Ta cap layer of the eb-deposited sample (Fig \ref{fig:4}a). This brighter contrast is also visible at the interface between the lower CoFe and the Ru layer of the rf-sputtered sample.

\begin{figure}
\centerline{
 \includegraphics[width=0.9\columnwidth]{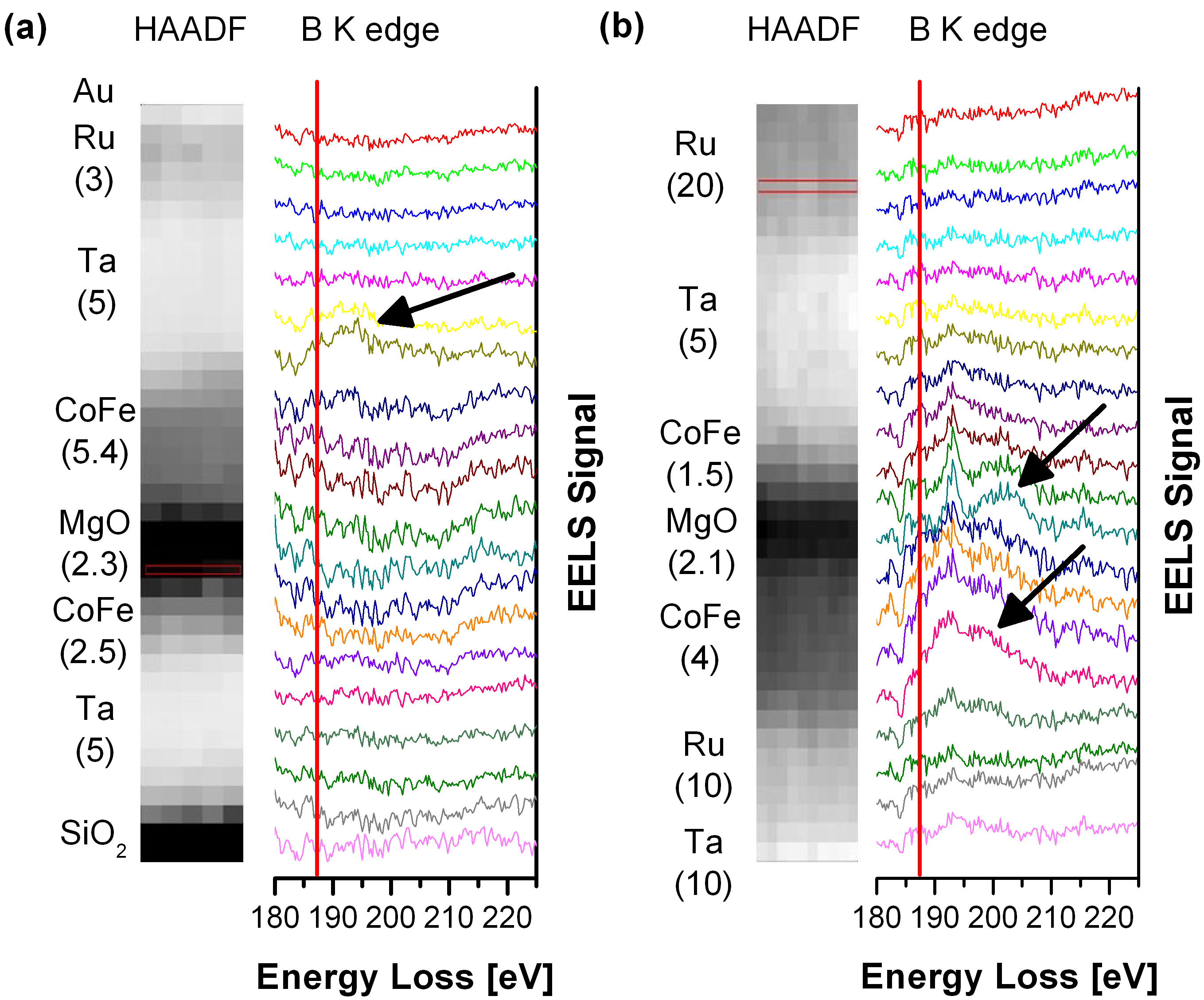}}
\caption{(color online) EEL spectra of MTJ with eb-deposited MgO barrier (a) shows absence of boron in the MgO and the MgO / CoFe interface and the presence within the CoFe / Ta interface region (arrow). In the rf-sputtered MTJ sample, boron is clearly visible in the MgO and the electrodes (arrows).}
\label{fig:5}
\end{figure}

{ Results of EELS investigations of the MTJs are summarized in Fig.~\ref{fig:5}. For the eb-deposited MgO barrier (Fig.~\ref{fig:5}a), results obtained from the model systems are confirmed: Boron is located in the interface region between the CoFe and the top Ta cap layer (see arrow) and not at the interface to the MgO again corroborating Ta to serve as a sink for boron. In contrast, boron signals are still detectable in both electrodes as well as in the MgO barrier in case of rf-sputter deposition (Fig.~\ref{fig:5}b). This indicates a higher density of those defects in MgO serving as diffusion vehicle for boron. 
}

\begin{figure}
\centerline{
 \includegraphics[width=1\columnwidth]{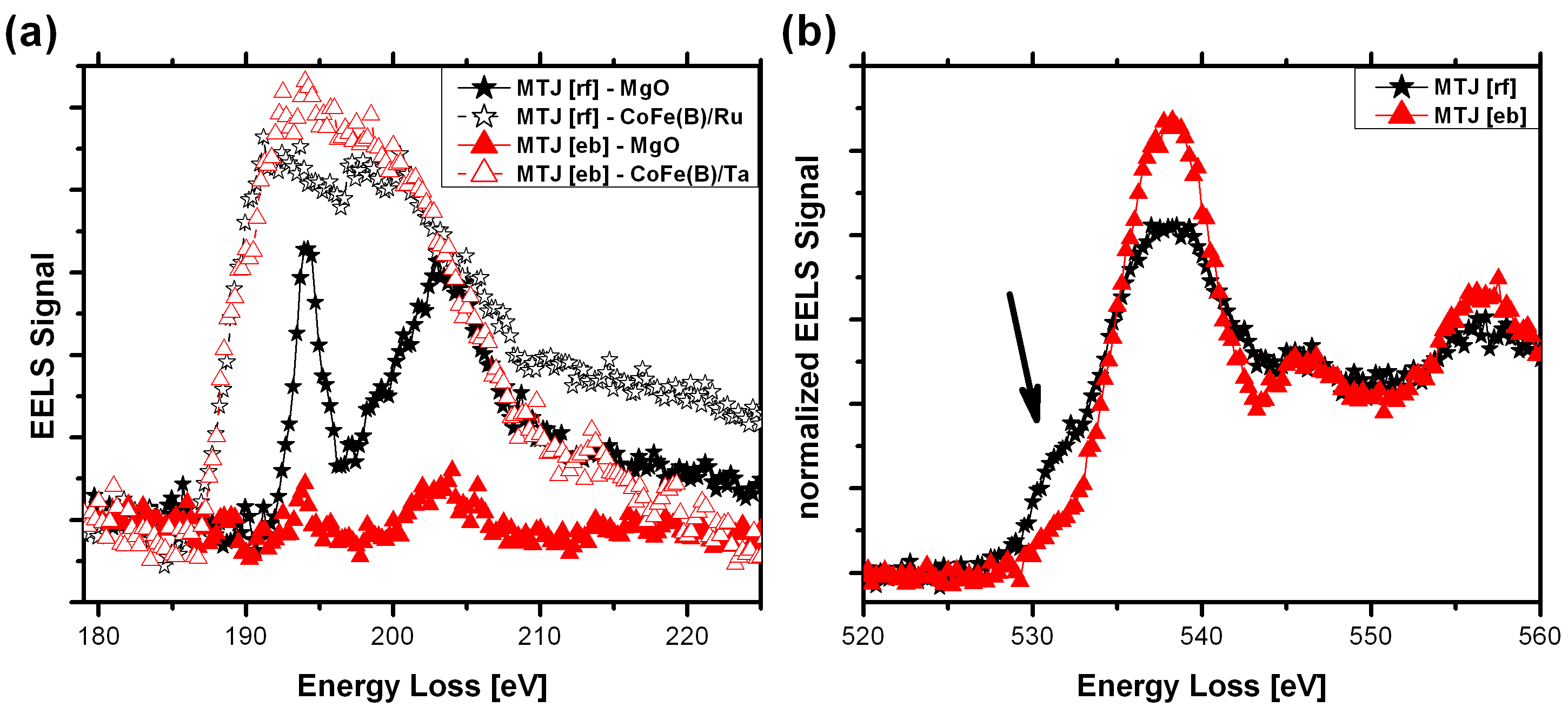}}
\caption{(color online) EELS absorption edges for e-beam and rf-sputtered magnetic tunnel junctions. (a) the B-K edge indicates metallic environment in electrode materials whereas the $\pi^*$ peak at 193 eV in the rf-sputtered MgO indicates oxidized boron in trigonal coordination with oxygen; please note the strongly reduced boron signal the e-beam deposited MgO. (b) the O-K edge shows two important features, i.e. sharper features in e-beam deposited MgO and a pre-peak structure (arrow) between 530 and 535 eV in the sputtered MgO indicating gap states\protect{\cite{Velev2007}}. }
\label{fig:6}
\end{figure}

Further insight into the effect of deposition technique on the boron and oxygen environment can be obtained from 
the (ELNES) fine structures at the B-K and O-K edges (Fig. \ref{fig:6}). Boron signals obtained in the CoFe as well as in the Ru and Ta cap layers indicate metallic bonding whereas the significant $\pi^*$ peak at 193 eV is characteristic for B in a [BO$_3$]$^{3-}$-like coordination with oxygen \cite{Sauer1993} (Fig.~\ref{fig:6}a).

Figure \ref{fig:6}b shows the O-K edge of the MgO barrier of both MTJ samples. Sharper features in the O-K edge are observed from the e-beam deposited MTJ sample. Following \cite{Plisch2001,Cha2007} indicating a less defective environment in the latter case 
since ELNES is rather sensitive to local variation of the atomic structure. In addition, a  pre-peak between 530 and 535 eV was observed in the O-K EELS signal of the rf-sputtered MTJ sample similar to that reported by Cha et al. \cite{Cha2007}. On the basis of first-principle calculations~\cite{Velev2007} those gap states have been related to band tailing of the conduction band of the MgO layer as a result of oxygen vacancies. These calculations also show that oxygen vacancies can significantly reduce the TMR from more than 1800\% for ideal systems to less than 800\%. In our case, TMR values of about 200\% have been measured for the e-beam deposited and rf-sputtered devices indicating that neither the enhanced boron concentration in the rf-sputtered MgO nor the observed gap states observed in the O-K-ELNES are limiting spin polarized transport. This might be related to the fact that, similar to results in \cite{Harnchana2013}, our EELS data are consistent with a uniform boron distribution in the MgO rather than accumulation at the barrier-electrode interfaces which is known to be detrimental for TMR\cite{Burton2006}.  These observations emphasize the crucial role of interfacial ordering for spin polarized transport and the importance of controlling segregation phenomena at CoFeB/MgO interfaces.

{ In summary, the present study shows that Ta cap layers serve as a sink for boron and do not interfere with the crystallization of amorphous CoFeB into crystalline CoFe which is intimately related to boron outdiffusion. Ru cap layers, however, act as boron diffusion barriers thus reducing the crystallization of CoFeB. In addition, polycrystalline Ru may act as a template for CoFeB crystallization which may lead to polycrystalline CoFe formation which we observed for thick CoFeB layers.  Additionally, e-beam deposited MgO barriers are free from any B or $\text{BO}_x$ contamination, in contrast to rf-sputtered MgO barriers which show a strong EELS signal typical for the [BO$_3$]$^{3-}$- environment after annealing in addition to gap states related to a pre-peak in the O-K-ELNES. Neither the boron content nor the gap states in rf-sputtered MgO affect spin polarized transport which is characterized by a TMR ratio of about 200\% for rf-sputtered and e-beam deposited MgO.}

This work was supported by the German Research Foundation (DFG) through SFB 602. Also, V.D. and A.T. would like to acknowledge the MIWF of the NRW state government for financial support. We gratefully acknowledge P. Peretzki for critical reading the manuscript.


\begin{thebibliography}{16}%
\makeatletter
\providecommand \@ifxundefined [1]{%
 \@ifx{#1\undefined}
}%
\providecommand \@ifnum [1]{%
 \ifnum #1\expandafter \@firstoftwo
 \else \expandafter \@secondoftwo
 \fi
}%
\providecommand \@ifx [1]{%
 \ifx #1\expandafter \@firstoftwo
 \else \expandafter \@secondoftwo
 \fi
}%
\providecommand \natexlab [1]{#1}%
\providecommand \enquote  [1]{``#1''}%
\providecommand \bibnamefont  [1]{#1}%
\providecommand \bibfnamefont [1]{#1}%
\providecommand \citenamefont [1]{#1}%
\providecommand \href@noop [0]{\@secondoftwo}%
\providecommand \href [0]{\begingroup \@sanitize@url \@href}%
\providecommand \@href[1]{\@@startlink{#1}\@@href}%
\providecommand \@@href[1]{\endgroup#1\@@endlink}%
\providecommand \@sanitize@url [0]{\catcode `\\12\catcode `\$12\catcode
  `\&12\catcode `\#12\catcode `\^12\catcode `\_12\catcode `\%12\relax}%
\providecommand \@@startlink[1]{}%
\providecommand \@@endlink[0]{}%
\providecommand \url  [0]{\begingroup\@sanitize@url \@url }%
\providecommand \@url [1]{\endgroup\@href {#1}{\urlprefix }}%
\providecommand \urlprefix  [0]{URL }%
\providecommand \Eprint [0]{\href }%
\@ifxundefined \urlstyle {%
  \providecommand \doi  [0]{\begingroup \@sanitize@url \@doi}%
  \providecommand \@doi [1]{\endgroup \@@startlink {\doibase
  #1}doi:\discretionary {}{}{}#1\@@endlink }%
}{%
  \providecommand \doi  [0]{doi:\discretionary{}{}{}\begingroup
  \urlstyle{rm}\Url }%
}%
\providecommand \doibase [0]{http://dx.doi.org/}%
\providecommand \Doi [0]{\begingroup \@sanitize@url \@Doi }%
\providecommand \@Doi  [1]{\endgroup\@@startlink{\doibase#1}\@@Doi}%
\providecommand \@@Doi [1]{#1\@@endlink}%
\providecommand \selectlanguage [0]{\@gobble}%
\providecommand \bibinfo  [0]{\@secondoftwo}%
\providecommand \bibfield  [0]{\@secondoftwo}%
\providecommand \translation [1]{[#1]}%
\providecommand \BibitemOpen [0]{}%
\providecommand \bibitemStop [0]{}%
\providecommand \bibitemNoStop [0]{.\EOS\space}%
\providecommand \EOS [0]{\spacefactor3000\relax}%
\providecommand \BibitemShut  [1]{\csname bibitem#1\endcsname}%
\bibitem [{\citenamefont {Miao}\ \emph {et~al.}(2011)\citenamefont {Miao},
  \citenamefont {M\"unzenberg},\ and\ \citenamefont {Moodera}}]{Miao2011}%
  \BibitemOpen
  \bibfield  {author} {\bibinfo {author} {\bibfnamefont {G.-X.}\ \bibnamefont
  {Miao}}, \bibinfo {author} {\bibfnamefont {M.}~\bibnamefont {M\"unzenberg}},
  \ and\ \bibinfo {author} {\bibfnamefont {J.~S.}\ \bibnamefont {Moodera}},\
  }\href {http://stacks.iop.org/0034-4885/74/i=3/a=036501} {\bibfield
  {journal} {\bibinfo  {journal} {Reports on Progress in Physics},\ }\textbf
  {\bibinfo {volume} {74}},\ \bibinfo {pages} {036501} (\bibinfo {year}
  {2011})}\BibitemShut {NoStop}%
\bibitem [{\citenamefont {Karthik}\ \emph {et~al.}(2012)\citenamefont
  {Karthik}, \citenamefont {Takahashi}, \citenamefont {Ohkubo}, \citenamefont
  {Hono}, \citenamefont {Gan}, \citenamefont {Ikeda},\ and\ \citenamefont
  {Ohno}}]{Karthik2012}%
  \BibitemOpen
  \bibfield  {author} {\bibinfo {author} {\bibfnamefont {S.~V.}\ \bibnamefont
  {Karthik}}, \bibinfo {author} {\bibfnamefont {Y.~K.}\ \bibnamefont
  {Takahashi}}, \bibinfo {author} {\bibfnamefont {T.}~\bibnamefont {Ohkubo}},
  \bibinfo {author} {\bibfnamefont {K.}~\bibnamefont {Hono}}, \bibinfo {author}
  {\bibfnamefont {H.~D.}\ \bibnamefont {Gan}}, \bibinfo {author} {\bibfnamefont
  {S.}~\bibnamefont {Ikeda}}, \ and\ \bibinfo {author} {\bibfnamefont
  {H.}~\bibnamefont {Ohno}},\ }\Doi {10.1063/1.4707964} {\bibfield  {journal}
  {\bibinfo  {journal} {Journal of Applied Physics},\ }\textbf {\bibinfo
  {volume} {111}},\ \bibinfo {pages} {083922 } (\bibinfo {year} {2012})},\ ISSN
  \bibinfo {issn} {0021-8979}\BibitemShut {NoStop}%
\bibitem [{\citenamefont {Thomas}\ \emph {et~al.}(2008)\citenamefont {Thomas},
  \citenamefont {Drewello}, \citenamefont {Sch\"afers}, \citenamefont
  {Weddemann}, \citenamefont {Reiss}, \citenamefont {Eilers}, \citenamefont
  {M\"unzenberg}, \citenamefont {Thiel},\ and\ \citenamefont
  {Seibt}}]{Thomas2008}%
  \BibitemOpen
  \bibfield  {author} {\bibinfo {author} {\bibfnamefont {A.}~\bibnamefont
  {Thomas}}, \bibinfo {author} {\bibfnamefont {V.}~\bibnamefont {Drewello}},
  \bibinfo {author} {\bibfnamefont {M.}~\bibnamefont {Sch\"afers}}, \bibinfo
  {author} {\bibfnamefont {A.}~\bibnamefont {Weddemann}}, \bibinfo {author}
  {\bibfnamefont {G.}~\bibnamefont {Reiss}}, \bibinfo {author} {\bibfnamefont
  {G.}~\bibnamefont {Eilers}}, \bibinfo {author} {\bibfnamefont
  {M.}~\bibnamefont {M\"unzenberg}}, \bibinfo {author} {\bibfnamefont
  {K.}~\bibnamefont {Thiel}}, \ and\ \bibinfo {author} {\bibfnamefont
  {M.}~\bibnamefont {Seibt}},\ }\Doi {10.1063/1.3001934} {\bibfield  {journal}
  {\bibinfo  {journal} {Applied Physics Letters},\ }\textbf {\bibinfo {volume}
  {93}},\ \bibinfo {eid} {152508} (\bibinfo {year} {2008})}\BibitemShut
  {NoStop}%
\bibitem [{\citenamefont {Sch\"afers}\ \emph {et~al.}(2009)\citenamefont
  {Sch\"afers}, \citenamefont {Drewello}, \citenamefont {Reiss}, \citenamefont
  {Thomas}, \citenamefont {Thiel}, \citenamefont {Eilers}, \citenamefont
  {M\"unzenberg}, \citenamefont {Schuhmann},\ and\ \citenamefont
  {Seibt}}]{Schafers2009}%
  \BibitemOpen
  \bibfield  {author} {\bibinfo {author} {\bibfnamefont {M.}~\bibnamefont
  {Sch\"afers}}, \bibinfo {author} {\bibfnamefont {V.}~\bibnamefont
  {Drewello}}, \bibinfo {author} {\bibfnamefont {G.}~\bibnamefont {Reiss}},
  \bibinfo {author} {\bibfnamefont {A.}~\bibnamefont {Thomas}}, \bibinfo
  {author} {\bibfnamefont {K.}~\bibnamefont {Thiel}}, \bibinfo {author}
  {\bibfnamefont {G.}~\bibnamefont {Eilers}}, \bibinfo {author} {\bibfnamefont
  {M.}~\bibnamefont {M\"unzenberg}}, \bibinfo {author} {\bibfnamefont
  {H.}~\bibnamefont {Schuhmann}}, \ and\ \bibinfo {author} {\bibfnamefont
  {M.}~\bibnamefont {Seibt}},\ }\Doi {10.1063/1.3272268} {\bibfield  {journal}
  {\bibinfo  {journal} {Applied Physics Letters},\ }\textbf {\bibinfo {volume}
  {95}},\ \bibinfo {eid} {232119} (\bibinfo {year} {2009})}\BibitemShut
  {NoStop}%
\bibitem [{\citenamefont {Eilers}\ \emph {et~al.}(2009)\citenamefont {Eilers},
  \citenamefont {Ulrichs}, \citenamefont {MÃ¼nzenberg}, \citenamefont {Thomas},
  \citenamefont {Thiel},\ and\ \citenamefont {Seibt}}]{Eilers2009b}%
  \BibitemOpen
  \bibfield  {author} {\bibinfo {author} {\bibfnamefont {G.}~\bibnamefont
  {Eilers}}, \bibinfo {author} {\bibfnamefont {H.}~\bibnamefont {Ulrichs}},
  \bibinfo {author} {\bibfnamefont {M.}~\bibnamefont {MÃ¼nzenberg}}, \bibinfo
  {author} {\bibfnamefont {A.}~\bibnamefont {Thomas}}, \bibinfo {author}
  {\bibfnamefont {K.}~\bibnamefont {Thiel}}, \ and\ \bibinfo {author}
  {\bibfnamefont {M.}~\bibnamefont {Seibt}},\ }\href@noop {} {\bibfield
  {journal} {\bibinfo  {journal} {J. Appl. Phys.},\ }\textbf {\bibinfo {volume}
  {105}} (\bibinfo {year} {2009})}\BibitemShut {NoStop}%
\bibitem [{\citenamefont {Burton}\ \emph {et~al.}(2006)\citenamefont {Burton},
  \citenamefont {Jaswal}, \citenamefont {Tsymbal}, \citenamefont {Mryasov},\
  and\ \citenamefont {Heinonen}}]{Burton2006}%
  \BibitemOpen
  \bibfield  {author} {\bibinfo {author} {\bibfnamefont {J.~D.}\ \bibnamefont
  {Burton}}, \bibinfo {author} {\bibfnamefont {S.~S.}\ \bibnamefont {Jaswal}},
  \bibinfo {author} {\bibfnamefont {E.~Y.}\ \bibnamefont {Tsymbal}}, \bibinfo
  {author} {\bibfnamefont {O.~N.}\ \bibnamefont {Mryasov}}, \ and\ \bibinfo
  {author} {\bibfnamefont {O.~G.}\ \bibnamefont {Heinonen}},\ }\Doi
  {10.1063/1.2360189} {\bibfield  {journal} {\bibinfo  {journal} {Applied
  Physics Letters},\ }\textbf {\bibinfo {volume} {89}},\ \bibinfo {eid}
  {142507} (\bibinfo {year} {2006})}\BibitemShut {NoStop}%
\bibitem [{\citenamefont {Heiliger}\ \emph {et~al.}(2007)\citenamefont
  {Heiliger}, \citenamefont {Gradhand}, \citenamefont {Zahn},\ and\
  \citenamefont {Mertig}}]{Heiliger2007}%
  \BibitemOpen
  \bibfield  {author} {\bibinfo {author} {\bibfnamefont {C.}~\bibnamefont
  {Heiliger}}, \bibinfo {author} {\bibfnamefont {M.}~\bibnamefont {Gradhand}},
  \bibinfo {author} {\bibfnamefont {P.}~\bibnamefont {Zahn}}, \ and\ \bibinfo
  {author} {\bibfnamefont {I.}~\bibnamefont {Mertig}},\ }\Doi
  {10.1103/PhysRevLett.99.066804} {\bibfield  {journal} {\bibinfo  {journal}
  {Physical Review Letters},\ }\textbf {\bibinfo {volume} {99}},\ \bibinfo
  {eid} {066804} (\bibinfo {year} {2007})}\BibitemShut {NoStop}%
\bibitem [{\citenamefont {Schreiber}\ \emph {et~al.}(2011)\citenamefont
  {Schreiber}, \citenamefont {suk Choi}, \citenamefont {Liu}, \citenamefont
  {Chiaramonti}, \citenamefont {Seidman},\ and\ \citenamefont
  {Petford-Long}}]{Schreiber2011}%
  \BibitemOpen
  \bibfield  {author} {\bibinfo {author} {\bibfnamefont {D.~K.}\ \bibnamefont
  {Schreiber}}, \bibinfo {author} {\bibfnamefont {Y.}~\bibnamefont {suk Choi}},
  \bibinfo {author} {\bibfnamefont {Y.}~\bibnamefont {Liu}}, \bibinfo {author}
  {\bibfnamefont {A.~N.}\ \bibnamefont {Chiaramonti}}, \bibinfo {author}
  {\bibfnamefont {D.~N.}\ \bibnamefont {Seidman}}, \ and\ \bibinfo {author}
  {\bibfnamefont {A.~K.}\ \bibnamefont {Petford-Long}},\ }\Doi
  {10.1063/1.3583569} {\bibfield  {journal} {\bibinfo  {journal} {Journal of
  Applied Physics},\ }\textbf {\bibinfo {volume} {109}},\ \bibinfo {eid}
  {103909} (\bibinfo {year} {2011})}\BibitemShut {NoStop}%
\bibitem [{\citenamefont {Cha}\ \emph {et~al.}(2009)\citenamefont {Cha},
  \citenamefont {Read}, \citenamefont {Egelhoff}, \citenamefont {Huang},
  \citenamefont {Tseng}, \citenamefont {Li}, \citenamefont {Buhrman},\ and\
  \citenamefont {Muller}}]{Cha2009}%
  \BibitemOpen
  \bibfield  {author} {\bibinfo {author} {\bibfnamefont {J.~J.}\ \bibnamefont
  {Cha}}, \bibinfo {author} {\bibfnamefont {J.~C.}\ \bibnamefont {Read}},
  \bibinfo {author} {\bibfnamefont {W.~F.}\ \bibnamefont {Egelhoff}}, \bibinfo
  {author} {\bibfnamefont {P.~Y.}\ \bibnamefont {Huang}}, \bibinfo {author}
  {\bibfnamefont {H.~W.}\ \bibnamefont {Tseng}}, \bibinfo {author}
  {\bibfnamefont {Y.}~\bibnamefont {Li}}, \bibinfo {author} {\bibfnamefont
  {R.~A.}\ \bibnamefont {Buhrman}}, \ and\ \bibinfo {author} {\bibfnamefont
  {D.~A.}\ \bibnamefont {Muller}},\ }\Doi {10.1063/1.3184766} {\bibfield
  {journal} {\bibinfo  {journal} {Applied Physics Letters},\ }\textbf {\bibinfo
  {volume} {95}},\ \bibinfo {pages} {032506 } (\bibinfo {year} {2009})},\ ISSN
  \bibinfo {issn} {0003-6951}\BibitemShut {NoStop}%
\bibitem [{\citenamefont {Karthik}\ \emph {et~al.}(2009)\citenamefont
  {Karthik}, \citenamefont {Takahashi}, \citenamefont {Ohkubo}, \citenamefont
  {Hono}, \citenamefont {Ikeda},\ and\ \citenamefont {Ohno}}]{Karthik2009}%
  \BibitemOpen
  \bibfield  {author} {\bibinfo {author} {\bibfnamefont {S.~V.}\ \bibnamefont
  {Karthik}}, \bibinfo {author} {\bibfnamefont {Y.~K.}\ \bibnamefont
  {Takahashi}}, \bibinfo {author} {\bibfnamefont {T.}~\bibnamefont {Ohkubo}},
  \bibinfo {author} {\bibfnamefont {K.}~\bibnamefont {Hono}}, \bibinfo {author}
  {\bibfnamefont {S.}~\bibnamefont {Ikeda}}, \ and\ \bibinfo {author}
  {\bibfnamefont {H.}~\bibnamefont {Ohno}},\ }\Doi {10.1063/1.3182817}
  {\bibfield  {journal} {\bibinfo  {journal} {Journal of Applied Physics},\
  }\textbf {\bibinfo {volume} {106}},\ \bibinfo {pages} {023920 } (\bibinfo
  {year} {2009})},\ ISSN \bibinfo {issn} {0021-8979}\BibitemShut {NoStop}%
\bibitem [{\citenamefont {Kurt}\ \emph {et~al.}(2010)\citenamefont {Kurt},
  \citenamefont {Rode}, \citenamefont {Oguz}, \citenamefont {Boese},
  \citenamefont {Faulkner},\ and\ \citenamefont {Coey}}]{Kurt2010}%
  \BibitemOpen
  \bibfield  {author} {\bibinfo {author} {\bibfnamefont {H.}~\bibnamefont
  {Kurt}}, \bibinfo {author} {\bibfnamefont {K.}~\bibnamefont {Rode}}, \bibinfo
  {author} {\bibfnamefont {K.}~\bibnamefont {Oguz}}, \bibinfo {author}
  {\bibfnamefont {M.}~\bibnamefont {Boese}}, \bibinfo {author} {\bibfnamefont
  {C.~C.}\ \bibnamefont {Faulkner}}, \ and\ \bibinfo {author} {\bibfnamefont
  {J.~M.~D.}\ \bibnamefont {Coey}},\ }\Doi {10.1063/1.3457475} {\bibfield
  {journal} {\bibinfo  {journal} {Applied Physics Letters},\ }\textbf {\bibinfo
  {volume} {96}},\ \bibinfo {pages} {262501 } (\bibinfo {year} {2010})},\ ISSN
  \bibinfo {issn} {0003-6951}\BibitemShut {NoStop}%
\bibitem [{\citenamefont {Cha}\ \emph {et~al.}(2007)\citenamefont {Cha},
  \citenamefont {Read}, \citenamefont {Buhrman},\ and\ \citenamefont
  {Muller}}]{Cha2007}%
  \BibitemOpen
  \bibfield  {author} {\bibinfo {author} {\bibfnamefont {J.~J.}\ \bibnamefont
  {Cha}}, \bibinfo {author} {\bibfnamefont {J.~C.}\ \bibnamefont {Read}},
  \bibinfo {author} {\bibfnamefont {R.~A.}\ \bibnamefont {Buhrman}}, \ and\
  \bibinfo {author} {\bibfnamefont {D.~A.}\ \bibnamefont {Muller}},\ }\Doi
  {10.1063/1.2769753} {\bibfield  {journal} {\bibinfo  {journal} {Applied
  Physics Letters},\ }\textbf {\bibinfo {volume} {91}},\ \bibinfo {eid}
  {062516} (\bibinfo {year} {2007})}\BibitemShut {NoStop}%
\bibitem [{\citenamefont {Sauer}\ \emph {et~al.}(1993)\citenamefont {Sauer},
  \citenamefont {Brydson}, \citenamefont {Rowley}, \citenamefont {Engel},\ and\
  \citenamefont {Thomas}}]{Sauer1993}%
  \BibitemOpen
  \bibfield  {author} {\bibinfo {author} {\bibfnamefont {H.}~\bibnamefont
  {Sauer}}, \bibinfo {author} {\bibfnamefont {R.}~\bibnamefont {Brydson}},
  \bibinfo {author} {\bibfnamefont {P.}~\bibnamefont {Rowley}}, \bibinfo
  {author} {\bibfnamefont {W.}~\bibnamefont {Engel}}, \ and\ \bibinfo {author}
  {\bibfnamefont {J.}~\bibnamefont {Thomas}},\ }\Doi
  {http://dx.doi.org/10.1016/0304-3991(93)90226-N} {\bibfield  {journal}
  {\bibinfo  {journal} {Ultramicroscopy},\ }\textbf {\bibinfo {volume} {49}},\
  \bibinfo {pages} {198 } (\bibinfo {year} {1993})},\ ISSN \bibinfo {issn}
  {0304-3991}\BibitemShut {NoStop}%
\bibitem [{\citenamefont {Plisch}\ \emph {et~al.}(2001)\citenamefont {Plisch},
  \citenamefont {Chang}, \citenamefont {Silcox},\ and\ \citenamefont
  {Buhrman}}]{Plisch2001}%
  \BibitemOpen
  \bibfield  {author} {\bibinfo {author} {\bibfnamefont {M.~J.}\ \bibnamefont
  {Plisch}}, \bibinfo {author} {\bibfnamefont {J.~L.}\ \bibnamefont {Chang}},
  \bibinfo {author} {\bibfnamefont {J.}~\bibnamefont {Silcox}}, \ and\ \bibinfo
  {author} {\bibfnamefont {R.~A.}\ \bibnamefont {Buhrman}},\ }\Doi
  {10.1063/1.1383569} {\bibfield  {journal} {\bibinfo  {journal} {Applied
  Physics Letters},\ }\textbf {\bibinfo {volume} {79}},\ \bibinfo {pages} {391}
  (\bibinfo {year} {2001})}\BibitemShut {NoStop}%
\bibitem [{\citenamefont {Velev}\ \emph {et~al.}(2007)\citenamefont {Velev},
  \citenamefont {Belashchenko}, \citenamefont {Jaswal},\ and\ \citenamefont
  {Tsymbal}}]{Velev2007}%
  \BibitemOpen
  \bibfield  {author} {\bibinfo {author} {\bibfnamefont {J.~P.}\ \bibnamefont
  {Velev}}, \bibinfo {author} {\bibfnamefont {K.~D.}\ \bibnamefont
  {Belashchenko}}, \bibinfo {author} {\bibfnamefont {S.~S.}\ \bibnamefont
  {Jaswal}}, \ and\ \bibinfo {author} {\bibfnamefont {E.~Y.}\ \bibnamefont
  {Tsymbal}},\ }\Doi {10.1063/1.2643027} {\bibfield  {journal} {\bibinfo
  {journal} {Applied Physics Letters},\ }\textbf {\bibinfo {volume} {90}},\
  \bibinfo {eid} {072502} (\bibinfo {year} {2007})}\BibitemShut {NoStop}%
\bibitem [{\citenamefont {Harnchana}\ \emph {et~al.}(2013)\citenamefont
  {Harnchana}, \citenamefont {Hindmarch}, \citenamefont {Sarahan},
  \citenamefont {Marrows}, \citenamefont {Brown},\ and\ \citenamefont
  {Brydson}}]{Harnchana2013}%
  \BibitemOpen
  \bibfield  {author} {\bibinfo {author} {\bibfnamefont {V.}~\bibnamefont
  {Harnchana}}, \bibinfo {author} {\bibfnamefont {A.~T.}\ \bibnamefont
  {Hindmarch}}, \bibinfo {author} {\bibfnamefont {M.~C.}\ \bibnamefont
  {Sarahan}}, \bibinfo {author} {\bibfnamefont {C.~H.}\ \bibnamefont
  {Marrows}}, \bibinfo {author} {\bibfnamefont {A.~P.}\ \bibnamefont {Brown}},
  \ and\ \bibinfo {author} {\bibfnamefont {R.~M.~D.}\ \bibnamefont {Brydson}},\
  }\Doi {http://dx.doi.org/10.1063/1.4802692} {\bibfield  {journal} {\bibinfo
  {journal} {Journal of Applied Physics},\ }\textbf {\bibinfo {volume} {113}},\
  \bibinfo {eid} {163502} (\bibinfo {year} {2013})}\BibitemShut {NoStop}%
\end{thebibliography}
%

\end{document}